# The Effect of Stereotypes on Perceived Competence of Indigenous Software Practitioners: A Professional Photo

Mary Sánchez-Gordón
*Department of Computer Science and Communication*
*Østfold University College*
Halden, Norway
mary.sanchez-gordon@hiof.no

Ricardo Colomo-Palacios
*Escuela Técnica Superior de Ingenieros Informáticos*
*Universidad Politécnica de Madrid*
Madrid, Spain
ricardo.colomo@upm.es

Cathy Guevara-Vega
*eCIER Research Group*
*Universidad Técnica del Norte*
Ibarra, Ecuador
cguevara@utn.edu.ec

Antonio Quiña-Mera
*eCIER Research Group*
*Universidad Técnica del Norte*
Ibarra, Ecuador
aquina@utn.edu.ec

*Abstract*— **Context: Potential employers can readily find job candidates' photos through various online sources such as former employers' websites or professional and social networks. The alignment or 'fit' between a candidate and an organization is inferred in online photos through dress style and presentations of self. On the other hand, for candidates from under-represented groups like Indigenous people traditional clothing is an important and lively aspect that allows them to express belonging, enter ceremony, and show resistance. Objective: This exploratory study aims to empirically demonstrate whether traditional clothing in a picture affects the evaluation of candidates' competence for a position like a software developer in which clothing should not be crucial. Method: We plan a quasi-experimental design with both candidates (photo models) and participants (evaluators) from IT companies. It follows a 2 x 2 x 2 design with dress style (traditional / non-traditional clothing), gender and race/ethnicity of the candidates as within-subjects factors. In addition, we will explore the evaluator's gender and experience in hiring as between-subjects factors.**

## I. Introduction

Software development is not only a technical and knowledge-intensive activity, but also a human-centric and collaborative one, that could benefit from the social attributes of the people involved in it [1]. In this scenario, large organizations such as Google and IBM established devoted offices of diversity and inclusion. These efforts have been successful in reducing explicit gender-based biases and discrimination towards minorities in the software development profession [2]. Diversity arises from demographic attributes that differentiate people (e.g., gender, age and ethnicity) or otherwise (e.g., role, expertise, personality) [3], [4].

Software developers have been identified as a special type of knowledge workers [5]. Software jobs involve mainly man–machine interactions requiring high levels of competence and education. When submitting job applications in countries where including a photo is a standard practice, like Ecuador, there are at least two factors that require consideration. Job candidates should consider not only what facial expression to adopt in the photo [6] but also what clothing to wear [7]. Studies examining how people perceive the same individuals when they adopt various facial expressions have also been conducted. According to [6], facial expression in a photo affects the perceived competence of candidates for the position of software developer regardless of evaluators' prior hiring experience for this type of job. Therefore, this factor should be considered.

Employers are increasingly using social media screening, also called cyber vetting, as part of their employment process [8]. It means that even in countries like the United States and Canada where attaching a photo to a job application is not recommended, pictures that the candidates have on the website of their previous employer or in online social networks like Instagram, Facebook, Twitter, and LinkedIn can still affect a hiring decision [6]. Previous research suggests that a job application can have different success rates depending on the candidate's picture regardless of whether the picture is in the CV or available on a social network [9]. Moreover, people can make judgments based on a photo after a minimal exposure time, of 100 ms [10]. As a result, people can draw specific trait inferences, e.g., trustworthiness most reliably, followed by competence.

Competence is one of the two fundamental dimensions of social cognition. The other dimension is warmth which reflects the survival need of knowing the intentions of others (positive or negative, perceived competition) whereas competence is the consequent ability to enact those intentions (status) [11]. From an evolutionary point of view, warmth is judged before competence and these dimensions not only impact impression formation but also underlie group stereotypes formed by combining high versus low levels of these two dimensions. According to Stereotype Content Model (SCM) [12], people perceive social groups and individuals based on how warm and competent they are. For instance, social groups are perceived as competent if they are high in status, e.g., educationally or economically successful [12]. In this sense, previous research has revealed that the race/ethnicity of people moderates such effects for Black Americans [13]. It is expected that this stereotype, called subtyping by class, plays a role in other groups who are not high in status like Andean Indigenous.

According to [8], hiring is not a value-free and neutral process in which the best candidate is selected. Hiring discrimination has been well documented based on social identities like gender, race, ethnicity, age, disability and religion [14]. In this context, stereotypes may affect the judgments of individual group members in one of two ways [15]. In an assimilative mode, stereotypes may lead individuals to judge group members consistently with the group's expectations. On the other hand, stereotypes may also produce contrastive effects. In our previous study [16] on factors influencing the software engineering career choice of Andean indigenous, the first mode, assimilative, was reported by most of the interviewees. They acknowledge the assimilative effects that stereotypes can have. Due to low expectations for Indigenous, equal performance from an Indigenous and non-Indigenous target is not sufficient to get a positive evaluation for Indigenous. Thus, more is needed from an Indigenous. That previous study [16] was, to the best of our knowledge, the first published that considers the effects of ethnicity diversity on software engineering career choice, and one of the very few studies of ethnicity diversity in software engineering [3], [4]. Software engineering research on diversity has not yet paid enough attention to ethnicity [17], [18].

Despite that studies have been conducted on the perceived competence of women [19], [20], little is known specifically about the degree to which stereotypes impact the perceived competence of software professionals from other under-represented groups like Indigenous. With an increasing indigenous population in Latin America, and Andean Indigenous professionals reporting discrimination in software industry [16], the effect of stereotypes on the perceived competence of indigenous software practitioners is a necessary category of study.

In their case, traditional clothing is an identity marker. In this sense, a recent study reveals that Andean Indigenous do not make evident their indigenousness by using non-traditional clothing and to some extent, avoid discrimination by blending in [21]. This is a potentially important gap in the literature, since traditional clothing uniquely identifies Andean indigenous, and provides non-indigenous with a ready way to classify indigenous as out-group members.

In this paper, we provide a new perspective of diversity in software development by investigating the effect of traditional Andean clothing on the perceived competence of Ecuadorian software developers. A survey was designed to better understand such an effect on these candidates. The contribution of this exploratory study is its high ecological validity since all the participants are software professionals from local IT companies who evaluate the competence of these candidates for a job of a software developer, specifically 24 faces of software professionals. Moreover, given that the majoritarian group in Ecuador is the "Mestizo" —mixed race—, half of the candidates will be Mestizos and the remaining will be Andean indigenous. We use the terms ethnicity and race interchangeably throughout this exploratory study since these terms are used to describe human diversity and they are irrevocably intertwined.

In the following section (2) we briefly present the research question and hypotheses. Then, participants are described (3) and the execution plan (4) is presented.

## II. RESEARCH QUESTION

We assume that cognitive bias impacts how potential employers perceive the candidates' competence. This exploratory study focuses on perceived competence of Indigenous software professionals. However, as we are not interested in how specific physiognomic differences —such as distance between eyes, shape of nose, height of forehead, width of face— affect the perception of individuals' competence, the focus is on the comparison of dress style (traditional Andean/non-traditional clothing) across the same set of 24 individuals. Therefore, this study formulates the following research question:

Does a choice of dress style (traditional Andean clothing vs non-traditional clothing) in a photograph influence the evaluation of an Ecuadorian software developer's competence in?

To answer this question, we plan to test the conceptual model illustrated in Fig. 1. It is based on established theories and previous studies related to stereotypes, the halo effect, and perceived competence.

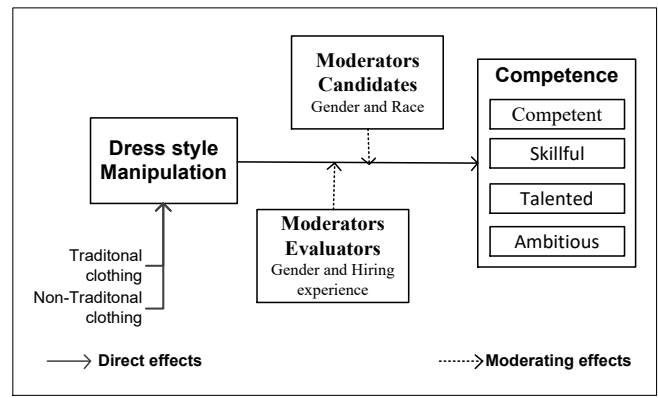

Fig. 1. A conceptual model that will be tested in this study

### A. Stereotype

Stereotypes often influence impression formation processes [12] and contribute to social biases, both positive and negative, which can lead to discrimination against individuals based on their social identities [22], [23]. The biggest categories are perceptions regarding gender, race/ethnicity, age, and nationality. These perceptions can trigger stereotypes, prejudices, or biases that one individual might hold towards another. The root of stereotypes explains why once stereotypes are formed, they are resistant to change [24]. However, it is worth noting that stereotype formation not only begins at a young age but also their use appears to be universal.

Stereotypes are constructs that tend to associate a set of traits and behaviors with a particular social group [22]. The societal stereotypes are also built from the perspective of dominant social groups, when analyzing stereotypes, it is necessary to account for the cultural and historical context [22], [23]. In Latin America, the majoritarian group is the "mestizo". For example, they are up to 72% of the Ecuadorian population [25] with high levels of Native American ancestry (up to 51%), followed by less European (up to 33%) and African (up to 13%) ancestry, according to a study about genetic data of a sample of mestizo population [26]. On the other hand, according to [27], discrimination based on phenotype —individual's appearance like height, hair color, skin color, and eye color— is an underemphasized aspect of race in the region, particularly in areas with a high indigenous population, like Ecuador. However, recent ethnographic studies have revealed the salience of phenotype in systems of racial classification in the region. It suggests that mestizos can share similar facial traits with Indigenous which result in a limited ability to distinguish between both groups.

Apart from gender discrimination, previous studies show that stereotypes against minorities in the tech industry based on race/ethnicity are persistent since there has been almost no change in racial equality and ethnic diversity in the industry [28]. At the core of systemic inequalities such as racism, the cognitive and affective responses of stereotypes and traits are present [22]. Oftentimes, the reasons for stereotypes and their inaccuracies are related to cumulative cultural evolution rooted in historical conditions [22], [23].

Latin America has a long history of discrimination against indigenous people from the beginning of the colonial period [29]. The declarations of independence of the region's republics mentioned the equality of rights of all citizens but did not alleviate the plight of indigenous peoples. According to the most recent census [30], in 2010, there were about 42 million indigenous people in Latin America representing nearly 8% of the total population. Although, it is difficult to estimate increases in the indigenous population across the region the only country that reported a decrease in its indigenous population in the past decade is Bolivia. A lower overall attainment of secondary and tertiary education in the indigenous population was also reported.

Sadly, even, if indigenous individuals achieve tertiary education, their earnings are often significantly lower than those of a non-indigenous person with the same qualifications [30]. In countries with large urban indigenous populations, such as Peru, Ecuador, Bolivia, and Mexico, the percentage of indigenous persons occupying high-skill jobs is consistently smaller than the percentage of non-indigenous people [30]. In Ecuador, they are about one-third as likely.

*B. The Halo Effect*

The "halo effect" is relevant to this study since it is a type of cognitive bias associated with stereotypes. The "halo effect" is generally defined as the impact of a global evaluation on evaluations of individual attributes of a person [31]. It occurs when one target is judged [32], therefore, global evaluations have the potential of altering perceptions of even relatively unambiguous stimuli about which an individual has sufficient information to render a confident judgment [31].

The halo effect has not been the main focus of previous software engineering studies. However, the talent assessment program at Gitlab includes best practices that help to avoid bias like the halo effect [33]. Moreover, it is also recognized that clothing and the halo effect are part of personal branding and help software developers to advance their developer careers [7].

Most psychological literature suggests that people draw inferences based on facial appearance judging mainly based on gender, age and ethnicity [34]. Previous research also suggests that traits that are considered attractive facially influence the level of trust and trusting behavior. This is due to the "halo effect" in which more "attractive people are perceived as good" [34], [35]. Additionally, stereotypes play a role in increasing both positive and negative perceptions of cognitive bias for social groups as shown in [34], [35] for the case of Muslims and hijab. In addition, studies suggest that this differential treatment caused by triggers like dress style leads to different outcomes in crucial aspects such as employment opportunities and occupational success [35].

Perceptions of individuals from different cultural backgrounds are also associated with specific cues that may reinforce stereotypes and influence trust. For example, in the case of Muslims, the main cues that activate negative perception are names and dress style [36]. Previous studies also show that clothing is a powerful non-verbal communication tool that activates stereotypes that can be either positive or negative [37], [38]. For example, grooming and dress style can influence evaluator appraisals of competence [38]. These biases can be extremely powerful in changing perceptions specially those based on race/ethnicity as individuals tend to have strong reactions to "stereotypical clothing" which is clothing associated with specific population groups [37]. In the case of Andean Indigenous people, they wear traditional clothing for many reasons, e.g., to express belonging, enter ceremony, and show resistance.

*C. Hyphoteses*

In our exploratory study, we focus on candidates that are Ecuadorian software developers who self-identify as Andean Indigenous or Mestizo. We control their age by focusing on a specific age range. Moreover, we control other two variables, nationality and cultural/historical context, by focusing on Ecuadorian software developers (for candidates) and IT companies located in Ecuador (for evaluators).

The null hypothesis (H0) states there is no difference in perceived competence between professional photos of job candidates wearing non-traditional and traditional clothing.

This study also explores the moderate effects of the following variables with these hypotheses:

H1: The gender and race of candidates can moderate the effect of their perceived competence.

H2: The gender and hiring experience of evaluators moderate the effect of candidates' perceived competence.

**Research design.** We plan a research design that manipulates the dress style related to the racial/ethnic group membership of a set of candidates. Each candidate is a photo model. We included two photos (experimental condition); in one of them candidates wear Andean traditional clothing and in the other, not. For the data collection, we create a survey that allows participants to evaluate the perceived competence based on candidates' photos. Thus, participants are evaluators in our design. Half of them will be randomly assigned to one of both experimental conditions, but without a pre-selection process. Therefore, the evaluation follows a quasi-experimental design. Participants will ask to evaluate 24 candidates —12 Mestizo and 12 Indigenous candidates, with an equal number of men and women in each group. Each candidate will be presented in predefine order on a separate page and participants will rate them on the competence dimensions.

The study follows a 2 (within-subjects, CandidateDress: traditional and non-traditional clothing) x 2 (within-subjects, CandidateRace: Mestizo and Indigenous) x 2 (within-subjects, CandidateGender: male and female) design. Moreover, we will explore the effect of evaluator's gender and experience in hiring.

**Ethical considerations.** This study involves human subjects; therefore, informed consent will be obtained from all participants. It describes the purpose and the main points of this study along with the potential (dis)advantages, a statement regarding privacy and confidentiality of their survey responses and information, and researchers' contact details. It is also relevant to define how research data will be treated (storage and accessibility) during and after the study's completion, therefore a Data Management Plan was also established.

We also checked that the study design will be carried out in line with the institutional guidelines of the universities involved in this study. It is worth noting that the ethical principles expressed in the Declaration of Helsinki guide this study. However, to check that this study complies with the local and regional legislations and standards, we will also submit it to the appropriate authorities of each university before beginning this study.

Finally, as this study includes photographs the models must provide informed consent for the use of these photographs in this study.

### III. Participants

We consider a convenience sample strategy to collect responses from employees of IT companies. We plan to contact companies located in the country of origin (Ecuador) from the candidates since cultural background influences the formation and development of stereotypes.

The number of participants in this study is not predetermined since it depends on how many employees of the IT companies involved in this study would complete the online questionnaire. However, we aim to collect a sample of around 200 software practitioners. To reduce the influence of social desirability bias, participation will be voluntary and anonymous. Cash incentives for participation will not be offered. We also declare that data will only be used for research purposes. Therefore, participants will not be motivated to purposefully misreport.

All participants who complete the questionnaire and have at least one year of experience in software development will be included in the analysis.

### IV. Execution Plan

At the core of our study, we experimentally test whether the use of traditional Andean/non-traditional clothing in photos can affect the perceived competence of job candidates in the Ecuadorian software industry. In addition, we aim to test the moderating roles of their gender and race as they can have an interaction on a positive or negative relationship; as well as whether the effects of the candidates' perceived competence depend on the gender and hiring experience of the evaluator (see Figure 1).

**Instrument design.** We created an anonymous questionnaire survey to facilitate self-reported data. Questions are grouped into two sections: basic information and evaluation.

The "basic demographic" section includes age, gender identity, education, job role and experience. Participant age data is collected using age ranges (e.g., 26–30 years) rather than specific ages. Gender identity has these options: man, woman, gender variant/non-conforming/non-binary, prefer to self-describe. For age and gender identity, the option "prefer not to say" is also included. Experience is measured in years.

The "evaluation of competence" section includes the evaluator's experience in hiring that is collected using a 3-point scale: frequently, occasionally, and never. For the measurement of perceived competence, previous research on the effect of three facial expressions on perception of competence of a software developer used a set of 20 pictures that was presented to 238 employees of IT companies [6]. Each candidate's competence was evaluated on a continuous rating scale by using 4-items proposed by Cuddy et al. [39] on how competent, skillful, talented, and ambitious the members of target groups are perceived to be. All the four items will load onto a single factor. Participants will be asked to rate the preselected target groups using a visual analog scale (VAS). This scale uses a horizontal line with textual anchors to represent a range of values, from "not at all" to "extremely". It is a continuous rating scale that allows more fine-grained answers [40]. Moreover, this type of scale has used in previous software engineering [41], [42]. A final question was included to ask participants if they knew any of the candidates. If so, we exclude the participant. In this way, we try to mitigate the bias.

**Stimuli creation.** All 24 photo models are employees of IT companies located in Ecuador. We included six Indigenous men, six Indigenous women, six Mestizo men, and six Mestizo women. To lessen the effect of age in hiring, we plan to select photo models within the same age range. An analysis of the public discourse about employability of software developers found that the most common definition of an old developer is 40+ years, followed by 30+ years and 35+ years [43]. As we are interested in an underrepresented group, we consider early adults between 22 and 34 years.

All models appeared in both experimental conditions (traditional Andean/non-traditional clothing) as it is the main target of the comparison. The use of traditional accessories will be restricted solely to traditional clothing to reduce possible biases. In this way, we aim to manipulate the dress style of the candidates. Moreover, the models will receive specific instructions regarding the type of facial expression (neutral) since we are only interested in gender and race comparisons and any other differences in the evaluation of different models will not be considered.

All pictures will be taken in the same place at a specific time of the day by the same camera digital with a maximum resolution of eight megapixels. Each picture is a facial photograph, head and shoulders of the models only, and portrait orientation, The place will be a room without a window to keep the light conditions constant. All models will receive an instruction to look straight at the camera and then adopt a neutral expression. Fig. 2 shows an example of stimulus material.

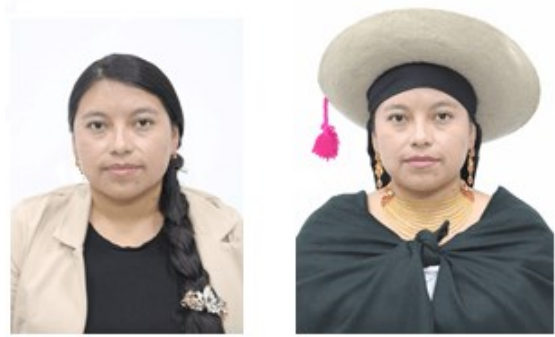

Fig. 2. Example of stimulus material: non-traditional/traditional clothing

**Pilot test.** We ask for feedback from two software developers. They analyze the question flow, check for any instances of confusion and register how long it takes them to complete the survey. Based on their feedback, a final version of the questionnaire will be created.

**Data collection.** We plan to invite IT companies located in Ecuador by phone. Once they agree to participate, an email will be sent to the contact person in the company and then, participants can access the survey online via a provided URL. Finally, the contact person will send us confirmation about participation.

Prior to participation in the online survey, participants will ask to review a consent document and "agree" to the contents of it. Then, participants will be randomly assigned across treatments —half of them to each one: treatment group and control group. Next, participants read that all the candidates they would see are equally qualified for the software developer job —having the required qualifications, education, and work experience. After that, the images of a total of 24 candidates are presented. The control group receives the candidates' photos wearing non-traditional clothing whereas the treatment group receives the candidates wearing traditional clothing.

The time to do each evaluation will be not limited so participants have as much time as needed to complete it. However, participants cannot fall back to a previous picture once they had moved on. Male faces will be presented on odd screens (e.g., first, third, fifth) while female faces were presented on even screens (e.g., second, fourth, sixth). Each of the two experimental groups will receive photos of the same models in the same order. A picture of the male Model 1, for instance, is always followed by a photo of the female Model 2, in turn, it is always followed by the picture of the male Model 3. In this way, we plan to keep the model sequence consistent to maintain all variables, except for the models' clothing, constant between the two experimental groups.

**Data Cleaning.** We will delete the consent form confirmation field since respondents could not continue without checking these boxes (the response is always "YES"). Moreover, participants who do not complete the survey will be removed. We will recode the raw data into a common quantitative coding scheme. For instance, 1 for never, 2 for occasionally, and 3 for frequently. The replication package will include the recoding instructions.

**Analysis.** We will provide descriptive statistics —frequency, means, and standard deviations— for the experimental factors.

A mixed model approach will be used since it considers both fixed effects (allowing to consider within and between group differences) and random effects which is an advantage of these type of models. Mixed models also provide the opportunity to analyze if needed only fixed, or only random, factors. Furthermore, using a mixed-model approach seems appropriate as our data represents a repeated-measurement design of both between-subjects (i.e., evaluator's experience in hiring) and within-subjects (i.e., race, gender, clothing manipulation) factors. This is also relevant since one advantage of mixed models over classical regression approaches is that one can control for correlations between measurements within subjects.

After determining the correct approach for our study, we have to check the model assumptions. Assumptions imply that our data must meet specific criteria for statistical method results to be accurate. The assumptions for mixed effects modeling include linearity, normality of the residuals, homogeneity of residual variance (homoscedasticity), no autocorrelation and no multicollinearity.

After assuring that the assumptions above are met, we plan to conduct the analysis. We will test the success of the photo clothing manipulation on the manipulation check and the dependent variable (i.e., competence). We also will test whether the effects of the candidates' perceived competence are moderated by their gender and race. In addition, we will test whether the effects of the candidates' perceived competence are moderated by the evaluators' gender and hiring experience. All models follow the same steps:

- The main effects of the predictors on the respective outcome variable are assessed. All variables will be tested to see if they have an individual statistical significance using a t-test.

- Two-way interactions among the predictor variables, the candidate's race and the candidate's gender are added. This will be done to explore whether the different interactions combinations between the two variables have an effect on our dependent variable. It is worth noting that if a specific variable at individual level showed no individual significance, sometimes when interactions are added it can have significance levels due to interrelations and overlapping.

- Three-way interactions between these variables are added.

The global significance level of our models will be tested by using an F test. Then, we will see the adjustment of our model through the goodness of fit checking the adjusted R squared value along with information criteria such as Akaike information criterion (AIK) and Hannan–Quinn (HQ).

Mixed models will be estimated to explore whether the effects of the manipulation differed depending on the indigenous self-identification (race/ethnicity) of the candidates.

Finally, we also plan to explore moderation by participants' gender and participants' hiring experience. It means we will conduct three-way interactions between the clothing manipulation, the candidate's gender, and the evaluators' gender (participants). In turn, we will conduct three-way interactions between the clothing manipulation, the candidate's gender, and the evaluators' hiring experience (participants).

**Validity Threats.** Both models and their evaluators are employees of IT companies. In this way, we can ensure ecological validity of the study. However, the quasi-experimental design has several limitations.

Regarding external validity, the statistical generalization of results is not possible since participants were not randomly selected from a population. Traditional clothing varies according to the Indigenous community, and some cannot be identified in a facial photograph, head and shoulders of the models only. Generalization to more complex real-life scenarios is limited and findings may not generalize to other types of populations.

Regarding construct validity, the perceived indigeneity is implemented as a single item, which may not accurately reflect fixation and different judges might produce different results. High internal validity is suggested by the controlled nature of the study. We also aim to ensure reproducibility by including the description of the experimental protocol and the replication package. Likewise, high conclusion validity is ensured by using well-understood statistical tests on data that meets their assumptions.

REFERENCES

[1] B. Vasilescu, V. Filkov, and A. Serebrenik, "Perceptions of Diversity on Git Hub: A User Survey," in *2015 IEEE/ACM 8th International Workshop on Cooperative and Human Aspects of Software Engineering*, May 2015, pp. 50–56. doi: 10.1109/CHASE.2015.14.

[2] Y. Wang and D. Redmiles, "Implicit Gender Biases in Professional Software Development: An Empirical Study," in *2019 IEEE/ACM 41st International Conference on Software Engineering: Software Engineering in Society (ICSE-SEIS)*, May 2019, pp. 1–10. doi: 10.1109/ICSE-SEIS.2019.00009.

[3] K. K. Silveira and R. Prikladnicki, "A Systematic Mapping Study of Diversity in Software Engineering: A Perspective from the Agile Methodologies," in *Proceedings of the 12th International Workshop on Cooperative and Human Aspects of Software Engineering*, in CHASE '19. Piscataway, NJ, USA: IEEE Press, 2019, pp. 7–10. doi: 10.1109/CHASE.2019.00010.

[4] Á. Menezes and R. Prikladnicki, "Diversity in Software Engineering," in *Proceedings of the 11th International Workshop on Cooperative and Human Aspects of Software Engineering*, in CHASE '18. New York, NY, USA: ACM, 2018, pp. 45–48. doi: 10.1145/3195836.3195857.

[5] A. N. Meyer, G. C. Murphy, T. Zimmermann, and T. Fritz, "Design Recommendations for Self-Monitoring in the Workplace: Studies in Software Development," *Proc. ACM Hum.-Comput. Interact.*, vol. 1, no. CSCW, p. 79:1-79:24, Dec. 2017, doi: 10.1145/3134714.

[6] P. Filkuková and M. Jørgensen, "How to pose for a professional photo: The effect of three facial expressions on perception of competence of a software developer," *Australian Journal of Psychology*, pp. 1–10, 2020, doi: 10.1111/ajpy.12285.

[7] Z. Nagy, "Your Online and Offline Presence," in *Soft Skills to Advance Your Developer Career: Actionable Steps to Help Maximize Your Potential*, Z. Nagy, Ed., Berkeley, CA: Apress, 2019, pp. 105–153. doi: 10.1007/978-1-4842-5092-1_4.

[8] J. Jacobson and A. Gruzd, "Cybervetting job applicants on social media: the new normal?," *Ethics Inf Technol*, vol. 22, no. 2, pp. 175–195, Jun. 2020, doi: 10.1007/s10676-020-09526-2.

[9] S. Baert, "Facebook profile picture appearance affects recruiters' first hiring decisions," *New Media & Society*, vol. 20, no. 3, pp. 1220–1239, Mar. 2018, doi: 10.1177/1461444816687294.

[10] J. Willis and A. Todorov, "First Impressions: Making Up Your Mind After a 100-Ms Exposure to a Face," *Psychological Science*, May 2016, Accessed: Aug. 13, 2020. [Online]. Available: https://journals.sagepub.com/doi/10.1111/j.1467-9280.2006.01750.x

[11] S. T. Fiske, A. J. C. Cuddy, and P. Glick, "Universal dimensions of social cognition: warmth and competence," *Trends in Cognitive Sciences*, vol. 11, no. 2, pp. 77–83, Feb. 2007, doi: 10.1016/j.tics.2006.11.005.

[12] S. T. Fiske, A. J. C. Cuddy, P. Glick, and J. Xu, "A model of (often mixed) stereotype content: Competence and warmth respectively follow from perceived status and competition," *Journal of Personality and Social Psychology*, vol. 82, pp. 878–902, 2002, doi: 10.1037/0022-3514.82.6.878.

[13] S. T. Fiske, H. B. Bergsieker, A. M. Russell, and L. Williams, "IMAGES OF BLACK AMERICANS: Then, 'Them,' and Now, 'Obama!,'" *Du Bois Review: Social Science Research on Race*, vol. 6, no. 1, pp. 83–101, ed 2009, doi: 10.1017/S1742058X0909002X.

[14] S. Baert, "Hiring Discrimination: An Overview of (Almost) All Correspondence Experiments Since 2005," in *Audit Studies: Behind the Scenes with Theory, Method, and Nuance*, S. M. Gaddis, Ed., in Methodos Series. Cham: Springer International Publishing, 2018, pp. 63–77. doi: 10.1007/978-3-319-71153-9_3.

[15] M. Biernat and D. Kobrynowicz, "Gender- and race-based standards of competence: lower minimum standards but higher ability standards for devalued groups," *J Pers Soc Psychol*, vol. 72, no. 3, pp. 544–557, Mar. 1997, doi: 10.1037//0022-3514.72.3.544.

[16] M. Sánchez-Gordón and R. Colomo-Palacios, "Factors influencing Software Engineering Career Choice of Andean Indigenous," in *Proceedings of the ACM/IEEE 42nd International Conference on Software Engineering: Companion Proceedings*, in ICSE '20. New York, NY, USA: ACM, Jun. 2020, pp. 264–265. doi: 10.1145/3377812.3390899.

[17] S. Harrison, "Five Years of Tech Diversity Reports—and Little Progress," *Wired*, 2019. Accessed: May 29, 2023. [Online]. Available: https://www.wired.com/story/five-years-tech-diversity-reports-little-progress/

[18] Y. A. Rankin and J. O. Thomas, "The Intersectional Experiences of Black Women in Computing," in *Proceedings of the 51st ACM Technical Symposium on Computer Science Education*, in SIGCSE '20. New York, NY, USA: Association for Computing Machinery, Feb. 2020, pp. 199–205. doi: 10.1145/3328778.3366873.

[19] N. Imtiaz, J. Middleton, J. Chakraborty, N. Robson, G. Bai, and E. Murphy-Hill, "Investigating the effects of gender bias on GitHub," in *Proceedings of the 41st International Conference on Software Engineering*, in ICSE '19. Montreal, Quebec, Canada: IEEE Press, May 2019, pp. 700–711. doi: 10.1109/ICSE.2019.00079.

[20] S. M. Hyrynsalmi, "The Underrepresentation of Women in the Software Industry: Thoughts from Career-Changing Women," in *2019 IEEE/ACM 2nd International Workshop on Gender Equality in Software Engineering (GE)*, May 2019, pp. 1–4. doi: 10.1109/GE.2019.00008.

[21] D. Román, D. Masaquiza, K. Ward, and L. Gonzalez-Quizhpe, "Contextualización transformativa de Educación Intercultural Bilingüe: A migrant Indigenous Andean community in the Galapagos Islands," *Journal of Multilingual and Multicultural Development*, vol. 0, no. 0, pp. 1–17, Feb. 2022, doi: 10.1080/01434632.2022.2036167.

[22] L. T. Harris, "Leveraging cultural narratives to promote trait inferences rather than stereotype activation during person perception," *Soc Personal Psychol Compass*, vol. 15, no. 6, Jun. 2021, doi: 10.1111/spc3.12598.

[23] D. Martin, S. J. Cunningham, J. Hutchison, G. Slessor, and K. Smith, "How societal stereotypes might form and evolve via cumulative cultural evolution," *Social and Personality Psychology Compass*, vol. 11, no. 9, p. e12338, Sep. 2017, doi: 10.1111/spc3.12338.

[24] A. Rattan and O. A. M. Georgeac, "Understanding intergroup relations through the lens of implicit theories (mindsets) of malleability," *Social and Personality Psychology Compass*, vol. 11, no. 4, p. e12305, Apr. 2017, doi: 10.1111/spc3.12305.

[25] Instituto Nacional de Estadística y Censos, "Resultados," *Instituto Nacional de Estadística y Censos*. https://www.ecuadorencifras.gob.ec/resultados/ (accessed May 28, 2023).

[26] N. Yang *et al.*, "Examination of ancestry and ethnic affiliation using highly informative diallelic DNA markers: application to diverse and admixed populations and implications for clinical epidemiology and forensic medicine," *Hum Genet*, vol. 118, no. 3–4, pp. 382–392, Dec. 2005, doi: 10.1007/s00439-005-0012-1.

[27] T. Ravindran, "The Power of Phenotype: Toward an Ethnography of Pigmentocracy in Andean Bolivia," *The Journal of Latin American and Caribbean Anthropology*, vol. 26, no. 2, pp. 219–236, 2021, doi: 10.1111/jlca.12551.

[28] S. Chattopadhyay, D. Ford, and T. Zimmermann, "Developers Who Vlog: Dismantling Stereotypes through Community and Identity," *Proc. ACM Hum.-Comput. Interact.*, vol. 5, no. CSCW2, pp. 1–33, Oct. 2021, doi: 10.1145/3479530.

[29] U. Nations, "Discrimination Against Indigenous Peoples: The Latin American Context," *United Nations*. https://www.un.org/en/chronicle/article/discrimination-against-indigenous-peoples-latin-american-context (accessed May 28, 2023).

[30] G. Freire *et al.*, "Indigenous Latin America in the twenty-first century : the first decade," The World Bank, 98544, Jan. 2015.

[31] R. E. Nisbett and T. D. Wilson, "The halo effect: Evidence for unconscious alteration of judgments," *Journal of Personality and Social Psychology*, vol. 35, pp. 250–256, 1977, doi: 10.1037/0022-3514.35.4.250.

[32] C. M. Judd, L. James-Hawkins, V. Yzerbyt, and Y. Kashima, "Fundamental dimensions of social judgment: understanding the relations between judgments of competence and warmth," *J Pers*

*Soc Psychol*, vol. 89, no. 6, pp. 899–913, Dec. 2005, doi: 10.1037/0022-3514.89.6.899.

[33] "Talent Assessment," *GitLab*. https://about.gitlab.com/handbook/people-group/talent-assessment/ (accessed May 27, 2023).

[34] M. Van Vugt and A. E. Grabo, "The Many Faces of Leadership: An Evolutionary-Psychology Approach," *Curr Dir Psychol Sci*, vol. 24, no. 6, pp. 484–489, Dec. 2015, doi: 10.1177/0963721415601971.

[35] S. Konrath and F. Handy, "The Good-looking Giver Effect: The Relationship Between Doing Good and Looking Good," *Nonprofit and Voluntary Sector Quarterly*, vol. 50, no. 2, pp. 283–311, Apr. 2021, doi: 10.1177/0899764020950835.

[36] C. Swank, "The Effect of Religious Dress on Perceived Attractiveness and Trustworthiness".

[37] N. J. Livingston and R. A. R. Gurung, "Trumping Racism: The Interactions of Stereotype Incongruent Clothing, Political Racial Rhetoric, and Prejudice Toward African Americans," *PsiChiJournal*, vol. 24, no. 1, pp. 52–60, 2019, doi: 10.24839/2325-7342.JN24.1.52.

[38] H. Wang, R. Zhang, L. Ding, and X. Mei, "Consistency matters: The interaction effect of grooming and dress style on hirability," *International Journal of Selection and Assessment*, vol. 30, no. 4, pp. 545–561, 2022, doi: 10.1111/ijsa.12399.

[39] A. J. C. Cuddy *et al.*, "Stereotype content model across cultures: towards universal similarities and some differences," *Br J Soc Psychol*, vol. 48, no. Pt 1, pp. 1–33, Mar. 2009, doi: 10.1348/014466608X314935.

[40] A. Voutilainen, T. Pitkäaho, T. Kvist, and K. Vehviläinen-Julkunen, "How to ask about patient satisfaction? The visual analogue scale is less vulnerable to confounding factors and ceiling effect than a symmetric Likert scale," *Journal of Advanced Nursing*, vol. 72, no. 4, pp. 946–957, 2016, doi: 10.1111/jan.12875.

[41] F. Ramin, "The role of egocentric bias in undergraduate Agile software development teams," in *Proceedings of the ACM/IEEE 42nd International Conference on Software Engineering: Companion Proceedings*, Seoul South Korea: ACM, Jun. 2020, pp. 122–124. doi: 10.1145/3377812.3382167.

[42] H. R. Neri and G. H. Travassos, "Software Quality is Multidimensional: Let's play with Tensors," in *Proceedings of the XXXIV Brazilian Symposium on Software Engineering*, Natal Brazil: ACM, Oct. 2020, pp. 126–131. doi: 10.1145/3422392.3422450.

[43] S. Baltes, G. Park, and A. Serebrenik, "Is 40 the new 60? How popular media portrays the employability of older software developers," *arXiv:2004.05847 [cs]*, Jun. 2020, [Online]. Available: http://arxiv.org/abs/2004.05847